\documentclass[a4paper,12pt]{article}
\usepackage{latexsym}
\usepackage{exscale}
\usepackage{graphicx}
\usepackage{amsmath}
\usepackage{amssymb}
\begin{document}
%
%
\title{Layer--resolved optical conductivity of $\mathrm{Co} \! \mid
\!\mathrm{Pt}$ multilayers}
\author{A.~Vernes$^{\, a)}$,
L.~Szunyogh$^{\, a,b)}$, L.~Udvardi$^{\, a,b)}$, P.~Weinberger$^{\, a)}$ \\
\ \\
$^{a)}$ {\small \it Center for Computational Materials Science}, \\
 {\small \it Technical University Vienna}, \\
 {\small \it Gumpendorferstr. 1a, 1060 Vienna, Austria} \\
$^{b)}$  {\small \it Department of Theoretical Physics }, \\
 {\small \it Budapest University of Technology and Economics} \\
 {\small \it Budafoki \'{u}t 8, 1521 Budapest, Hungary}}
\renewcommand{\today}{}
\maketitle
%
%
%
%
\begin{abstract}
The complex optical conductivity tensor is calculated for the
$\mathrm{Co} \! \mid \!\mathrm{Pt}$ multilayer systems by applying a
contour integration technique within the framework of the
spin--polarized relativistic screened Korringa--Kohn--Rostoker method.
It is shown that the optical conductivity of the $\mathrm{Co} \! \mid
\!\mathrm{Pt}$ multilayer systems is dominated by contributions
arising from the Pt cap and/or substrate layers.
\end{abstract}
%
%
%
%
{\it Contact author}: A. Vernes, Center for Computational Materials
Science, \\ Technical University Vienna, Gumpendorferstr. 1a, A--1060
Vienna, Austria; \\ 
Fax: +43--1--58801--15898; E-mail: av@cms.tuwien.ac.at .
%
\section{Introduction}
\label{sect:intro}
%

Since $\mathrm{Co} \! \mid \!\mathrm{Pt}$ multilayer systems can be
used as magneto--optical recording media \cite{HTB97}, in the last
decade a large amount of experimental investigations has been
performed in these systems; realistic theoretical investigations of
these multilayer systems, however, are still lacking.

The contour integration technique \cite{SW99} permits the evaluation
of the complex optical conductivity tensor as given by the Kubo
formula at nonzero temperatures and for finite life--time broadening.
This technique used together with the spin--polarized relativistic
screened Korringa--Kohn--Rostoker (SKKR) method \cite{SUWK94} provides
then the most proper description of magneto--optical properties in
layered systems by accounting on the same footing for both the inter--
and the intra--band contributions \cite{VSW01b}.

>From a numerical standpoint of view, the computational accuracy is
permanently controlled by using recently developed algorithms such as
the cumulative special--points method \cite{VSW01a}.  All results
given in the present contribution have been obtained with an accuracy
of 0.001 a.u. by using 35 (2) Matsubara poles at 300 K in the upper
(lower) semi--plane and a life--time broadening of 0.048 Ryd. The
Fermi level -0.038 Ryd corresponds to that of fcc--Pt bulk (lattice
parameter of 7.4137 a.u.).

%
\section{Results and discussions}
\label{sect:results}
%

Experimentally it has been shown that Pt as a substrate promotes an
fcc(111) texture \cite{WBG+92}. For this reason here only the results
obtained for this particular surface orientation are presented.

\begin{figure}[hbtp] \centering
\begin{tabular}{cc}
\includegraphics[width=0.47\columnwidth,clip]{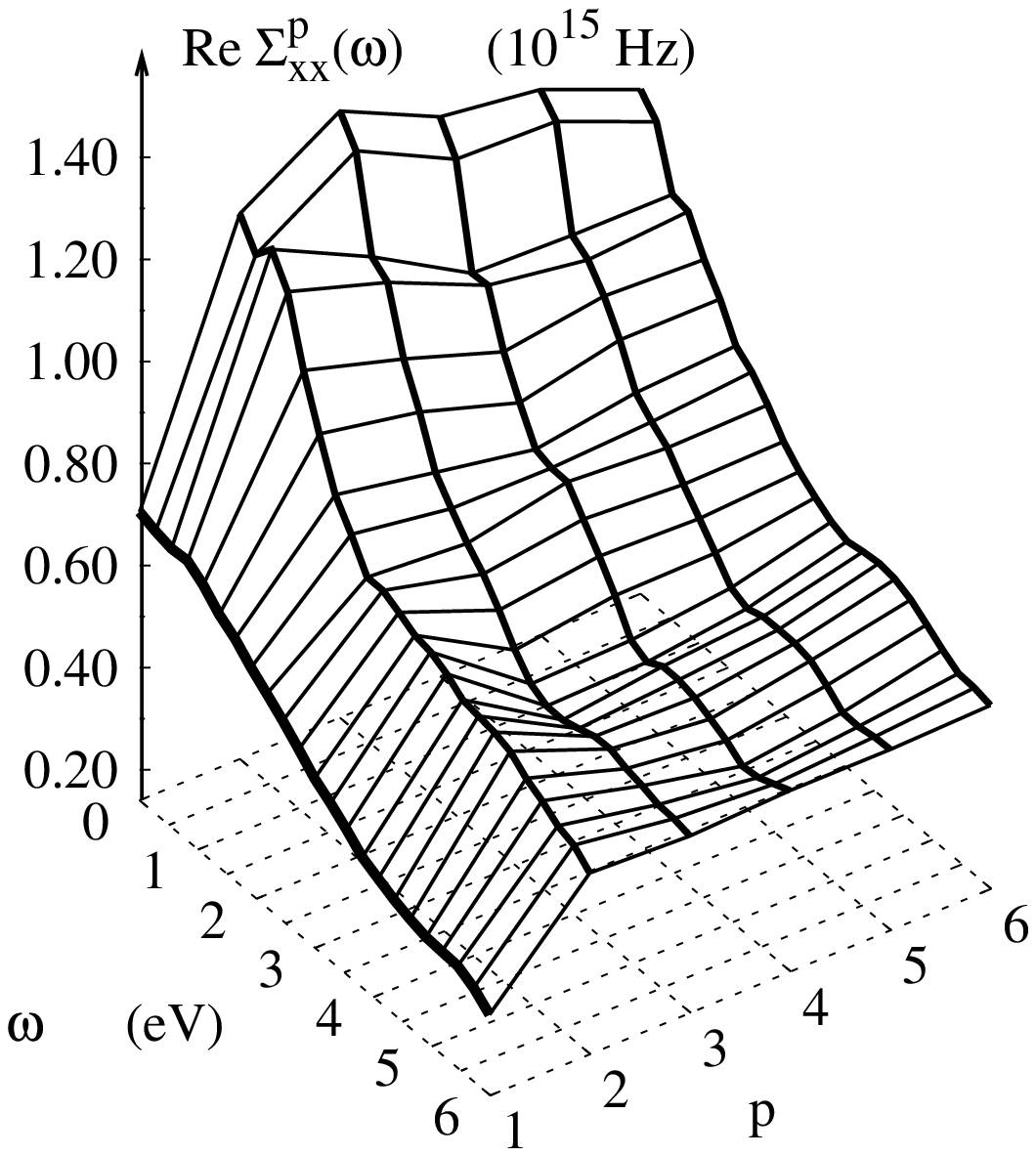} &
\includegraphics[width=0.47\columnwidth,clip]{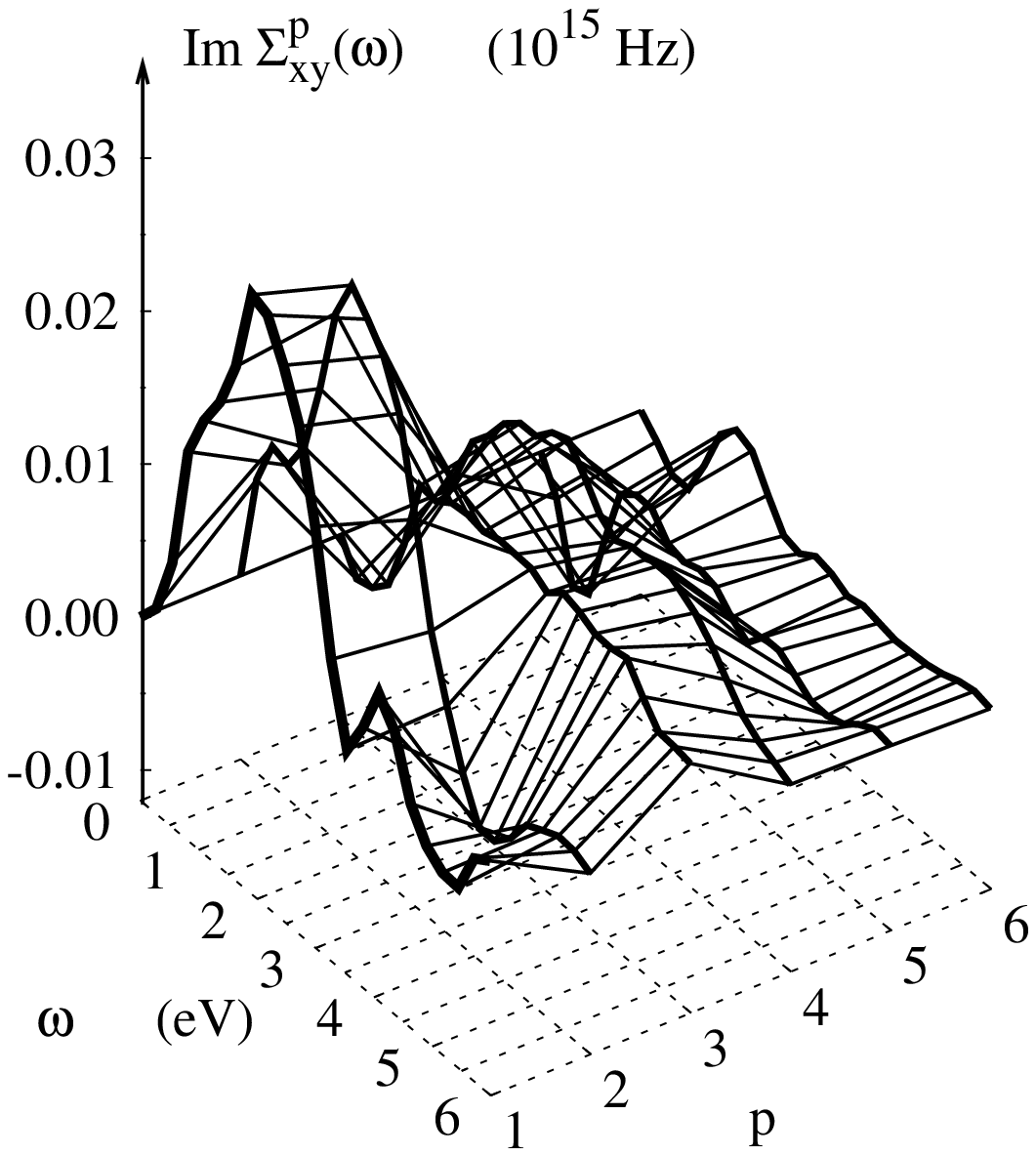} 
\end{tabular}
\caption[
      Absorptive parts of the layer--resolved complex optical
      conductivity $\protect\Sigma^{\,p}_{\mu\nu}(\omega)$ for
      fcc(111)--$\protect\mathrm{Co}\!\mid\!\mathrm{Pt}_{5}$ as a function of the
      optical frequency $\protect\omega$ and the layer index $p$. Heavy line marks the
      Co--layer resolved optical conductivity
      $\protect\Sigma^{\,p=1}_{\mu\nu}(\omega)$.
]
    {\label{fig:sgm-copt}
}
\end{figure}

As can be seen from Fig.\ \ref{fig:sgm-copt}, in the case of
fcc(111)--$\mathrm{Co} \! \mid \! \mathrm{Pt}_{5}$ -- over the whole
range of optical frequencies $\omega$ -- the real part of the complex
optical conductivity $\Sigma_{\mathrm{xx}}(\omega)$ is mainly
determinated by contributions arising from the Pt substrate layers.
>From all these Pt contributions, the largest arises from the first
layer below the surface Co--layer.  From 0 eV up to 3.5 eV the
contribution of the first Pt--layer below the surface to the imaginary
part of $\Sigma_{\mathrm{xy}}(\omega)$ almost equals that of the
Co--layer.  Around 4 eV, the Co contribution to $\mathrm{Im}\
\Sigma_{\mathrm{xy}}(\omega)$ has a local maximum and the first
Pt--layer a minimum. The other Pt substrate layers contribute
more or less in--between these two extrema. The other parts of the
complex optical conductivity tensor, i.e., $\mathrm{Im}\
\Sigma_{\mathrm{xx}}(\omega)$ and $\mathrm{Re}\
\Sigma_{\mathrm{xy}}(\omega)$ show similarities with those given in
Fig.\ \ref{fig:sgm-copt}: $\mathrm{Im}\ \Sigma_{\mathrm{xx}}(\omega)$
is dominated by the contribution from the first Pt--layer and
$\mathrm{Re}\ \Sigma_{\mathrm{xy}}(\omega)$ is almost equally
determinated by the surface Co--layer and by the first Pt substrate
layer.  Furthermore, based on our calculations performed for fcc(100)
and fcc(110) surface orientations, it can be said that the above
listed features of layer--resolved optical conductivities are not
surface dependent.

\begin{figure}[hbtp] \centering
\begin{tabular}{cc}
\includegraphics[width=0.47\columnwidth,clip]{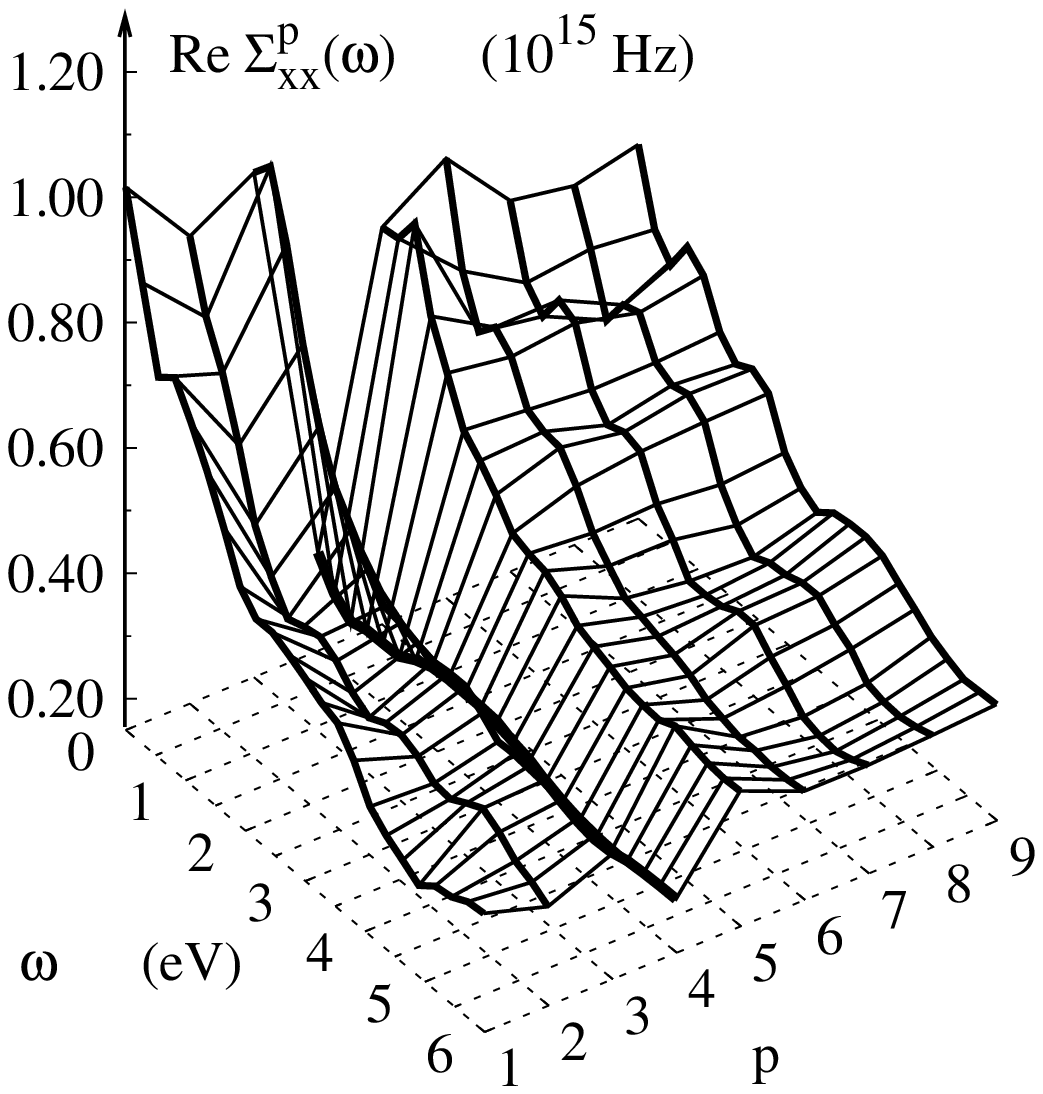} &
\includegraphics[width=0.47\columnwidth,clip]{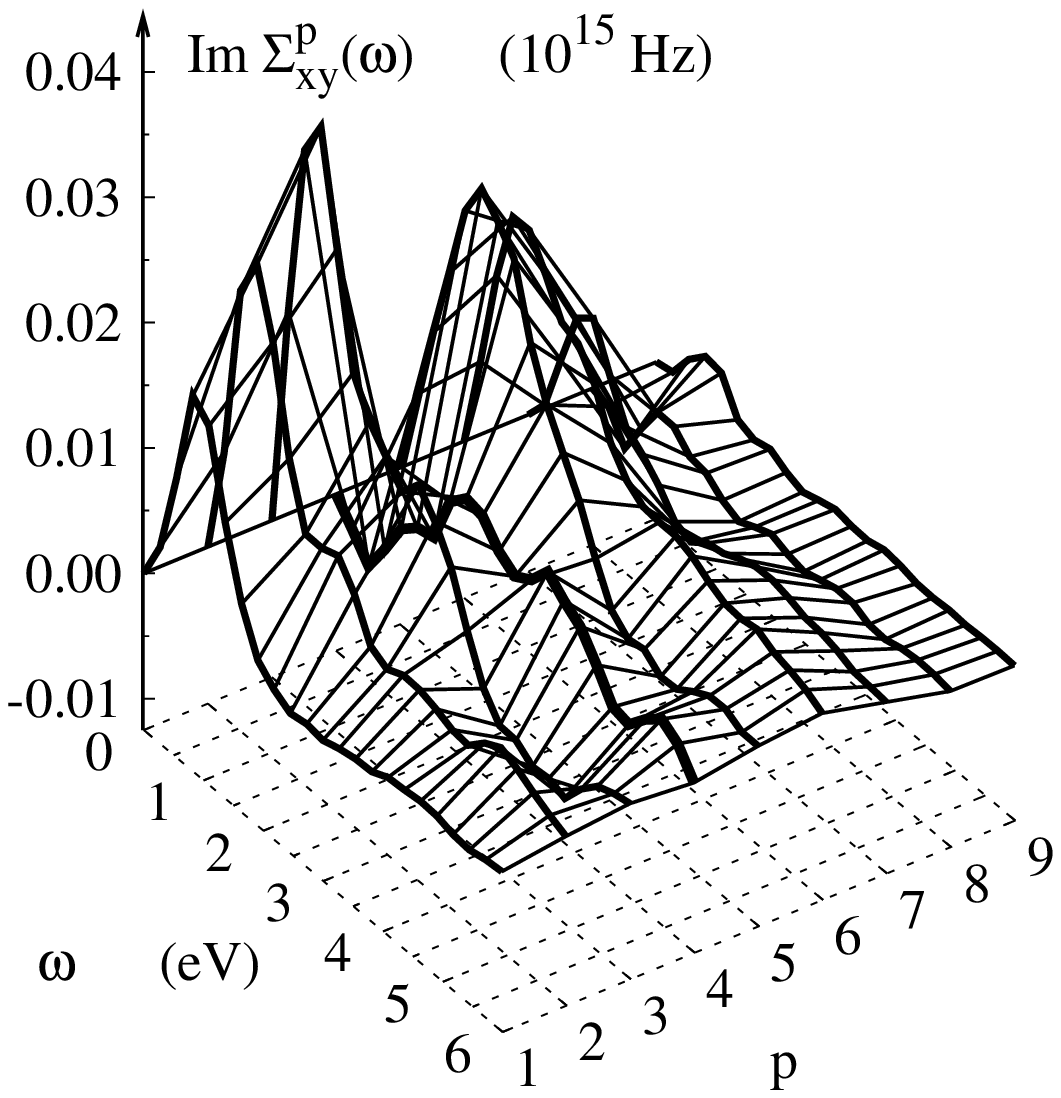} 
\end{tabular}
\caption[
      As in Fig.\ 1 for
      fcc(111)--$\protect\mathrm{Pt}_{3}\!\mid\!\mathrm{Co}\!\mid\!\mathrm{Pt}_{5}$.
]
    {\label{fig:sgm-ptcopt}
}
\end{figure}

Pt cap layers on the top of the Co--layer deposited on fcc(111)--Pt
are needed to prevent oxidation of the surface \cite{FBT00}.  We found
that only multilayer systems with Pt cap layers exhibit perpendicular
magnetization to the surface.  Furthermore, it has been found that the
magnetic moments induced in the Pt--layers are symmetrically
distributed above and below the Co--layer in such systems. Therefore,
it is not surprising that a cap and the corresponding Pt substrate
layer contribute in a similar manner to the layer--resolved optical
conductivity tensor as is demonstrated in Fig.\
\ref{fig:sgm-ptcopt}. Additional calculations performed for
fcc(111)--$\mathrm{Pt}_{m} \! \mid \!  \mathrm{Co} \!  \mid \!
\mathrm{Pt}_{8-m}$ for $m=1,2$ (not shown in Fig.\
\ref{fig:sgm-ptcopt}), proved that this particular feature remains
unchanged when increasing the number of Pt cap layers.  A comparison
of Fig.\ \ref{fig:sgm-copt} with Fig.\ \ref{fig:sgm-ptcopt} shows that
the Pt cap layers reduce the contributions from the Co and the Pt
substrate layers, but do not change significantly their frequency
dependence.

In conclusion, it can be said that over the whole range of frequencies
besides the ferromagnetic Co--layer the optical conductivity tensor of
$\mathrm{Co} \! \mid \!\mathrm{Pt}$ multilayer systems is equally
determinated by Pt contributions arising from a Pt cap and/or
substrate layers.
%
\section{Acknowledgements}
%
This work was supported by the Austrian Ministry of Science (Contract
No. 45.451), by the Hungarian National Science Foundation (Contract
No. OTKA T030240 and T029813) and partially by the RTN network
``Computational Magnetoelectronics'' (Contract
No. HPRN--CT--2000-00143).
%
%

%
%
\newpage
\listoffigures
\newpage
%
 
%
\newpage
%
 
%
\end{document}